\documentclass[twocolumn]{revtex4}
\usepackage{epsf}
\usepackage{graphicx}           

\begin{document}          


\title{Optimizing momentum space DMRG using quantum information entropy}

\author{\"O.~Legeza$^{(1,2)}$ and J. S{\'o}lyom$^{(1)}$}


\affiliation{$^{(1)}$Research Institute for Solid State Physics, H-1525 Budapest, P.\ O.\ Box 49, Hungary \\ 
$^{(2)}$Chair of Theoretical Chemistry, Friedrich--Alexander University Erlangen--Nuremberg. D-91058 Erlangen, Egerlandstr. 3, Germany }

\date{\today}

\vskip -8pt
\begin{abstract}
In order to optimize the ordering of the lattice sites in the momentum 
space and quantum chemistry versions of the density matrix renormalization group (DMRG)  
method we have studied the separability and entanglement of the target state for the 
1-D Hubbard model and various molecules. 
By analyzing the behavior of von Neumann 
and Neumann-Renyi entropies we have found criteria that 
help to fasten convergence. 
A new initialization procedure has been developed which maximizes
the Kullback-Leibler entropy and 
extends the active space (AS) in a dynamical fashion. The dynamically extended active space
(DEAS) procedure 
reduces significantly the effective system size during the first half sweep and accelerates the speed
of convergence of momentum space DMRG and quantum chemistry DMRG to a great extent.  
The effect of lattice site ordering on the number of block states to be kept during the RG procedure
is also investigated.
\end{abstract}

\pacs{PACS number: 75.10.Jm}

\maketitle

\section{Introduction}

The density matrix renormalization group (DMRG) method 
\cite{white1,white2} has been widely used in recent years to study coupled fermionic 
and spin chain problems. Its application 
got a new momentum during the past few years when it was reformulated to models
defined in momentum space \cite{xiang} (MS-DMRG) or to quantum chemistry calculations  
\cite{white3,white4} (QC-DMRG).
The properties of the Hubbard model \cite{erick1} and small diatomic molecules 
\cite{mitrushenkov,chan,legeza1,legeza2} 
have been studied using these methods.  
 However, several new technical problems are 
raised by these versions of the DMRG that have to be solved 
to increase the efficiency of the method and to stabilize 
its performance. 

A main difference between MS-DMRG and homogeneous lattice models studied by standard real space DMRG 
with periodic boundary condition is that in the
latter case each lattice site is equivalent and carries the same amount of information.  
In contrast to this, k-points or molecular orbitals lying closer to or farther 
away from the Fermi surface have different information content. The method is very sensitive to 
the ordering of the k-points
or molecular orbitals and an optimal ordering would have a major impact on the performance 
of MS-DMRG \cite{erick1,chan,legeza1,legeza2}. In fact 
the method can loose the target state 
and converge to a local minimum if an inappropriate ordering is used.
It has also been found using the dynamical block state selection (DBSS) approach \cite{legeza1} 
that the same accuracy can be achieved with more
or less block states, depending on the ordering.

The density matrices of composite 
systems, the separability of states and the nature of entanglement
have been extensively studied 
\cite{barnum1,barnum2,horodecki1,horodecki2,zyczkowski1,jozsa5,vidal1,vidal2,braunstein1}
in the past few years.  
Since MS-DMRG represents a composite system with long-range interactions the
results of quantum information theory  
can be used to understand the criteria of convergence of MS-DMRG.

Another major feature which hindered the powerful application of MS-DMRG is the 
lack of the so called infinite lattice method which in the real space version
generates a relatively accurate starting configuration for the so called 
finite lattice method.
In the initialization procedure proposed by Xiang \cite{xiang} the environment blocks of various lengths
were generated in advance of the finite lattice method. It is expected that a better procedure 
can be developed to generate the environment blocks by taking into account the 
change of the system block states during the process of renormalization.
This is also crucial for 
reaching faster the crossover between the environment error and the truncation error.
Then using the DBSS approach the accuracy  
is controlled by the threshold value of  
the truncation error fixed in advance of the calculations. 
The new procedure relies on finding the most important states carrying the largest information. The
space of these states will be called active space (AS) following a similar notation in quantum chemistry,
namely the CAS of the 
complete active space self consistent field (CASSCF) method 
used to define the orbitals for the CI treatment. As it has been shown long ago
\cite{bauschlicher} in the multi reference configuration interaction (MRCI)
calculations 
the convergence depends not only on the size of the active space but on other
constraints of the numerical treatment as well.          
It is, therefore, a 
very important task to develop a protocol
that extends the active space more effectively.
 
From the point of view of synergetics DMRG can be interpreted as a dynamical system. 
In this analogy 
the response of a given model system to incident messages is studied and  
the change of the relative importance of messages is determined 
as the method converges to an attractor.  
Therefore, besides the practical importance to  
improve the MS-DMRG procedure, 
the study of the interaction
of the subsystem blocks of DMRG as a function of the ordering of lattice sites 
is a very interesting question  
from the point of view of information theory, synergetics and quantum data compression 
\cite{schumacher1,jozsa1,jozsa2,jozsa3}.
 
Our aim in this paper is 1) to study the criteria of convergence of MS-DMRG by analyzing
the structure of the superblock and subsystem density matrices and 
quantities used in quantum 
information theory, 2)   
to develop  
a more efficient initialization procedure that 
collects the most important block states required to describe the total system in a better way
and 3) to study the effect of ordering on quantum data compression.

The setup of the paper is as follows. In Sec.~II we describe the theoretical background of  
separability and entanglement, mutual entropy and 
relative importance of incident messages, and 
generation of information entropy. Numerical investigation of these quantities in the context of
MS-DMRG and QC-DMRG is presented in Sec.~III and optimization of ordering is shown in Sec.~IV. 
Sec.~V. is devoted to the main steps of a new protocol that uses a dynamically extended active space 
to improve the initialization procedure.  
It is shown how the Abelian point-group 
symmetry can be used in the framework of QC-DMRG.     
Numerical results obtained for the 1-D Hubbard model and for various molecules are presented in Sec.~VI. and
the effect of ordering on quantum compression is also analyzed in some detail. 
The summary of 
our conclusions and future prospectives are presented in Sec.~VII.

\section{Theoretical background}

\subsection{Separability versus entanglement}

In general, if a finite system is divided into smaller subsystems (blocks)  
the Hamiltonian of the finite system is constructed from  
terms acting inside the blocks and the interaction terms among 
the blocks. The Hilbert space ($\cal G$) of dimension $N$ of the system is formed from the direct 
product states of the subsystem basis states. In particular, if $\cal G$ 
describes a composite system with $m$ subsystems then,
\begin{equation}
{\cal G} = \otimes_{i=1}^m {\cal G}_i,\,\,\,
\dim {\cal G} = \prod_{i=1}^m N_i = N.
\end{equation}
In general, states of ${\cal G}$ can be pure states or mixed states described
by the density matrix written as   
\begin{equation}
\rho = \sum_i p_i|\psi_i\rangle\langle\psi_i|,
\end{equation}
where $|\psi_i\rangle$ are eigenstates of the Hamiltonian acting on ${\cal G}$ and $\sum_i p_i = 1$.
The density matrix $\rho$ has the following properties: (i) ${\rm Tr} \rho=1$,
(ii) $\rho$ is a positive operator, i.e., ${\rm Tr}(\rho P)\geq 0$ for any projector $P$,
(iii) $\rho$ can be represented by its spectral decomposition as
\begin{equation}
\rho = \sum_{n=1}^N \omega_n P_n, \,\,\, \sum_{n=1}^N \omega_n=1,\,\,\, \omega_n \geq 0,
\end{equation}
where $P_n$ form a complete set of orthogonal projectors. 
A state $\rho$ is called separable if it can be 
written in the form 
\begin{equation}
\rho = \sum_{i} p_i \otimes_{l=1}^m \rho_i^l,
\end{equation}
where $\rho_i^l$ are states on ${\cal G}_l$, thus the subsystems are either not correlated or 
their correlation is purely classical.

\begin{figure}
\includegraphics[scale=0.35]{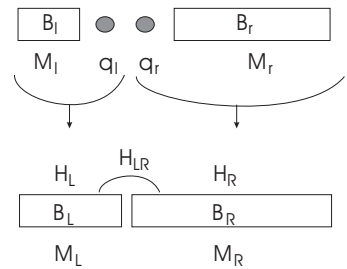}
\vskip .4cm
\caption{Schematic plot of system and environment block of DMRG and related quantities described in the text.}
\label{fig:dmrg}
\end{figure}

In the density matrix renormalization group
method proposed by S. R. White \cite{white1,white2} 
the total system called the superblock 
has two sites with
$q_l$ and $q_r$ degrees of freedom 
between the left and right blocks
$B_l$ and $B_r$ of dimensions $M_l$ and $M_r$, respectively.
Thus the superblock 
can be considered as a 
specially constructed composite system. 
Its configuration is shown in Fig.~\ref{fig:dmrg}. 

The so called target state ($|\psi_T\rangle$) of the superblock system which is the  
eigenstate that one wants to calculate is formed from the direct product states of the 
blocks and the sites. It can be a 
coherent or incoherent superposition of several eigenstates. In the first case one deals with
a pure state, while the latter corresponds to a mixed state. 
In this paper we examine only
the first case, i.e., $|\psi_T\rangle$ is chosen to be a pure state.

If we combine the $B_l \bullet$ composite system to one subsystem and  
$\bullet B_r$ to another one, then the so called superblock Hamiltonian for such a bi-partite system
(in this paper bi-partite system refers to a composite system with $m=2$ and not the conventional
term used in solid state physics to denote a crystal with specially arranged sublattices)  
consists of   
interaction terms determined in the blocks, denoted as ${\cal H}_L$ and ${\cal H}_R$ 
and interaction terms between the blocks denoted as ${\cal H}_{LR}$.
The relative contribution of each term to the superblock energy can be measured as 
\begin{equation}
\langle {\cal H}_i \rangle = \frac{\langle\Psi_T|{\cal H}_i|\Psi_T\rangle}{\langle\Psi_T|\sum_i {{\cal H}_i}|\Psi_T\rangle},
\label{eq:hexp}
\end{equation} 
where $i\equiv L,R,LR$ and $\sum_i \langle {\cal H}_i \rangle = \langle \sum_i {\cal H}_i \rangle$.

Since the target state is a pure state it follows from the   
Schmidt decomposition that for    
$|\Psi_T \rangle \in {\cal G} = {\cal G}_L\otimes {\cal G}_R$, with $\dim {\cal G}_L=M_L$,
 $\dim {\cal G}_R=M_R$, $M_L \times M_R = N$,  
\begin{equation}
|\Psi_T \rangle = \sum_{i=1}^{r\leq\min(M_L,M_R)} \omega_i |e_i\rangle \otimes |f_i\rangle, 
\label{eq:schmidt}
\end{equation}
where $|e_i \rangle \otimes |f_i \rangle $ form a bi-orthogonal basis 
$\langle e_i | e_j \rangle = \langle f_i | f_j \rangle = \delta_{ij}$, and $0\leq \omega_i \leq 1$ 
with the condition $\sum_i \omega_i^2 = 1$. 
If $r>1$ and 
in the range of a state $\rho$ there exists a $|\Psi\rangle$ such that
\begin{eqnarray}
\Lambda  & = & \langle \Psi | \rho^{-1} | \Psi \rangle ^{-1} > \frac{1}{1+\max_{i\neq j}(\omega_i\omega_j)},
\label{eq:sepcond1}
\\ 
\Lambda  & = & \langle \Psi | \rho | \Psi \rangle > \max_i \omega_i^2, 
\label{eq:sepcond}
\end{eqnarray}
then according to Ref.~[\onlinecite{zyczkowski1}] $\rho$ is inseparable. 
If $\rho$ has an eigenvector $|\Psi\rangle$ corresponding to the eigenvalue $\Lambda$ 
such that the conditions in Eqs.~(\ref{eq:sepcond1})-(\ref{eq:sepcond}) hold, then $\rho$ is separable.
In our
case for a pure state $|\Psi\rangle=|\Psi_T\rangle$ and $\Lambda=1$. Using singular value decomposition to generate 
the Schmidt coefficients one can easily check if the conditions of Eqs.~(\ref{eq:sepcond1})-(\ref{eq:sepcond}) hold
or not for a given target state.

Even if the target state is a pure state the subsystems can be in a mixed state
$\rho_L$ and $\rho_R$. 
A quantitative characterization of the degree of mixtures is provided by the von Neumann
entropy 
\begin{equation}
S(\rho) = -{\rm Tr}(\rho \ln \rho) 
\label{eq:neumann}
\end{equation}
and the so called participation number 
\begin{equation}
R(\rho)=\frac{1}{{\rm Tr}(\rho^2)}\,.
\label{eq:partip}
\end{equation}
The Neumann entropy is zero for a pure state and 
$S=\ln N$ for a totally mixed states
with $\rho_{I} = \frac{1}{N}I$ where $I$ denotes the identity matrix of dimension $N$.
The participation number 
varies from unity for pure states to $N$ to the totally mixed states 
and can be interpreted as an 
effective number of the states in the mixture. The Renyi entropy
$S_q(\rho)=(\ln({\rm Tr}\rho^q))/(1-q)$ with $q>1$ can also be used to measure how much a given state is mixed.
The necessary conditions for a density matrix to be separable have been worked
out by Peres \cite{peres1} and Horodecki {\sl et al} \cite{horodecki6}.

A fundamental concept related to inseparability and  
non-locality of quantum mechanics is the entanglement.  
A typical example of the maximally entangled pure state 
for two spin-1/2 particles is
\begin{equation}
|\psi\rangle = \frac{1}{\sqrt 2} |\uparrow \downarrow\rangle \pm |\downarrow \uparrow\rangle. 
\label{eq:entang}
\end{equation}
In order to measure the degree of entanglement between the blocks of DMRG one can make use of 
the entanglement monotone defined as 
$E:\rho({\cal G}_L\otimes{\cal G}_R)\rightarrow R_+$ with (i) $E(\rho)=0$ if $\rho$ is separable, (ii) $E$ is 
convex and (iii) $E$ is non-increasing (on average) under 
LOCC (local quantum operations (quantum operations on the left or right block) or classical communications).
A particular entanglement monotone, the entanglement of formation \cite{bennett2,wootters1,vedral2} is defined as 
\begin{equation}
E_F(\rho) \equiv \min \sum_i p_i S({\rm Tr_L}|\psi_i\rangle\langle\psi_i|),
\label{eq:entform}
\end{equation}
where the minimum is taken over all the possible realizations of the state $\rho$, 
${\rm Tr_L}$ is a partial trace with respect to the right block, $S$ is the Neumann entropy defined 
in Eq.~(\ref{eq:neumann}). For a pure state Eq.~(\ref{eq:entform}) can be easily calculated. 
A mixed state is entangled if it cannot be represented as a mixture of factorizable pure states and
there has been a great effort to determine a measure of entanglement for mixed states of a bipartite 
system \cite{vedral2,bennett1,lewenstein1}.
In fact, according to Bennett \cite{bennett3} all inseparable mixed states have non-zero 
"entanglement of formation" which means that a non-zero 
amount of pure entangled states are needed to build them. 

\subsection{Mutual entropy and Kullback-Leibler entropy}

\begin{figure}
\includegraphics[scale=0.4]{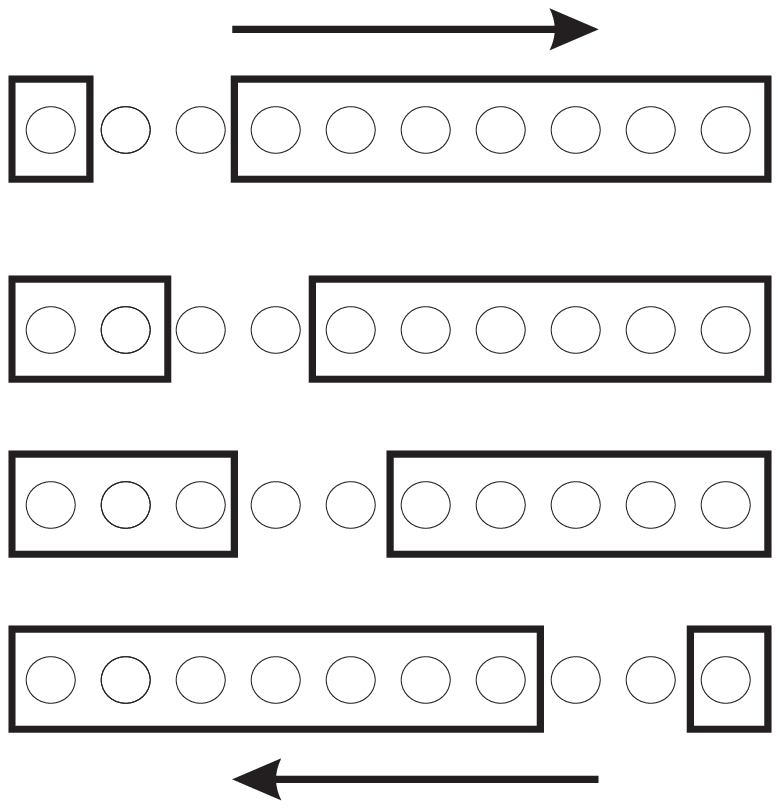}
\vskip .2 cm
\caption{The decomposition of the finite system to subsystems as a function of iteration steps corresponding to various partitionings.}
\label{fig:fin01}
\end{figure}

In the DMRG procedure the complete Hilbert spaces of the blocks are truncated, thus they
generate only a restricted subspace of the total Hilbert space. The various
partitionings of the finite system containing $L$ lattice sites 
to subsystem blocks are obtained by systematically changing the sizes of
the left and right blocks ($l$) and ($r$), respectively, with $l+2+r=L$ as shown in Fig.~\ref{fig:fin01}. 
In our implementation, described later in the text, the first iteration step corresponds to $l=1$ and $r=L-2-l$ and 
$l$ is increased as long as $l=L-3$.
The renormalization
procedure is used to obtain better configurations of the
block states for a given partitioning yielding a more accurate
description of the total system. In the forward sweep the left and right blocks are called the
system and environment block, respectively. In the backward sweep the right block becomes the system block
and the left block the environment block.
In the DMRG method, 
the renormalized states of the  
system block are selected from the eigenstates of the reduced density 
matrix of the system block having the largest eigenvalues. The reduced density matrix is formed from the 
target state as 
$\rho_L={\rm Tr_R} \rho$ and $\rho_R = {\rm Tr_L} \rho$ for the left and right subsystems, respectively, and
$\rho = |\psi_T \rangle \langle \psi_T|$ is  
determined by the diagonalization of the superblock Hamiltonian.
It is thus evident that since the various possible partitionings of the system  
have a strong effect on the structure of ${\cal H}_L, {\cal H}_R, {\cal H}_{LR}$ and on the density matrix 
of the superblock system, 
the structure of the reduced density matrix of the system block also depends on the environment block to 
a great extent. 
In this respect, states with  
largest eigenvalues of the reduced density matrix of the system block can be 
considered as dominant states while states with smaller weights 
as recessive states. The term recessive is used to indicate that such a block 
state gives no considerable contribution to the superblock wave function with 
that given environment block, however, it is possible that it will provide a 
considerable contribution when it interacts with an environment block in a subsequent sweep  
of the finite lattice algorithm. The $M_{min}$ 
parameter introduced in Ref.~[\onlinecite{legeza1}] 
ensures that these recessive states are also carried on during 
the sweeping procedure until they might  
become a dominant state.

For a given system block various environment blocks can be constructed and  
analogously to the genetic algorithm we can treat them as different species of a population
with different information content. 
Alternatively, we can think of the environment blocks 
as sources of messages. A meaning can be attributed to a message if the response 
of the receiver, in our case the system block, is taken into
account. Thus one can define the relative importance of message $\omega_j$ as the eigenvalues
of the reduced density matrix. In other words, even if the environment block 
contains all the states of the restricted Hilbert space defined on the $r=L-l-2$ sites of
the environment block, 
its information content can be very small if the system block is truncated so much  
that after taking into account the conservation of quantum numbers the Hilbert space of the
total system (superblock) is reduced drastically. The opposite treatment when the 
system block is considered as the 
source of messages and the environment block as a receiver works in the same way. 
Therefore, one needs a protocol to measure the conditional (mutual) information content of the blocks.  

Before proceeding further we recall a few definitions from
classical information theory in order to describe Neumann entropy and its connection to 
quantum information theory. Let us assume that there are two events (not independent in general) 
described by two ensembles $(X^L, X^R)$
and two sets of probability distributions of the elements denoted as $p(x^L)$ and $p(x^R)$.
We have used the labels $L$ and $R$ to indicate that the two events will later be related to the 
DMRG blocks. The
Shannon entropy \cite{shannon} for the two sets is defined as 
\begin{equation}
S_i = -\sum_{j=1}^{M_i} p(x^i_j) \ln p(x^i_j), \,\, \sum_j p(x^i_j) = 1
\label{eq:entropy}
\end{equation}    
where $i\equiv L,R$. 
It is worth to note that the entropy depends only on the probability distribution, $p(x_j^i)$.
The largest uncertainty of an event corresponds to the uniform distribution 
with $p(x^i_j)=1/M_i$ where $M_i$ is the number of elements. If  
the state of the event is known exactly, then  
$S=0$ since $p(x^i_j) = 1$ and all $p(x_{j^\prime \neq j})=0$. 
Following the notation of Haken \cite{haken} the expression Eq.~(\ref{eq:entropy}) can be interpreted as an 
average of the quantity over $f_j$  
\begin{equation}
S = \sum_j p(x_j) f_j, \,\, 
\end{equation}    
where $f_j = - \ln p(x_j)$ and the weight is $p(x_j)$.
In this respect $f_j$ is the information content of the state with 
index $j$ and $p(x_j)$ is the probability or relative frequency. 
If under a different condition  
we find a new relative frequency
$p(x_j^\prime)$ then the change of information is 
\begin{equation}
\Delta_j = \ln p(x^\prime_j) - \ln p(x_j).
\label{eq:delta_j}
\end{equation}    
To obtain the mean change of information, we average Eq.~(\ref{eq:delta_j}) over the new distribution function 
$p(x^\prime_j)$ and obtain the so called Kullback information gain as  
\begin{equation}
K(p(x^\prime),p(x)) = \sum_j p(x_j^\prime) \Delta_j = \sum_j p(x_j^\prime) \ln \frac{p(x_j^\prime)} {p(x_j)}
\label{eq:kullback}
\end{equation}    
where $\sum_j p(x_j) = 1$ and $\sum_j p(x_j^\prime) = 1$.
Eq.~(\ref{eq:kullback}) has an important property, namely  
\begin{equation}
K(p(x^\prime), p(x)) \geq 0.
\end{equation}    
If $X^L$ and $X^R$ are not independent of each other then 
the so called mutual information $I(L,R)$ quantifies the correlation between the two events, i.e., it gives 
the information about $X^L$ provided $X^R$ is known. $I(L,R)$ is written as
\begin{equation}
\begin{array}{rcl}
I(L,R) & = & S_L - S(L|R)  \\ 
       & = & S_L + S_R - S_{LR}  \\
       & = & S_R - S(R|L),
\label{eq:information}
\end{array}
\end{equation}
where the conditional entropy $S(L|R)$ is given by 
\begin{equation}
\begin{array}{rcl}
S(L|R) & = & S_{LR} - S_R \\
       & = & - \sum_{j,j^\prime} p(x_j^L, x_{j^\prime}^R) \ln p(x_j^L, x_{j^\prime}^R) \\ 
       &   & + \sum_{j} p(x_j^R) \ln p(x_j^R)  
\end{array}
\end{equation}    
and the total entropy $S_{LR}$ is calculated from the joint probability distribution of the two events 
$p(x_j^L, x_{j^\prime}^R)$.  
It is clear from Eq.~(\ref{eq:information}) that $I(L,R)$ is symmetric under the interchange of $X^L$ and $X^R$
and zero if and only if the two events are completely uncorrelated, i.e., when $p(x^L, x^R) = p(x^L)p(x^R)$.

The quantum analogue of the Shannon entropy is
the Neumann entropy. In the context of DMRG it can be defined 
 with $p(x_j^i)=\omega_j^i$ where $\omega_j^i$ are the eigenvalues of the reduced
density matrices of the subsystems. 
The two events can be related to the left and right subsystem blocks. 
Since the target state was chosen to be a pure state it corresponds to zero Neumann entropy 
with $S_{LR} = 0$, thus 
$I(L,R) = S_L + S_R$. Recalling the Schmidt decomposition of Eq.~(\ref{eq:schmidt}) based
on singular value decomposition one easily obtains that 
$\rho_L$ and $\rho_R$ have the same set of non-zero eigenvalues
($\omega_i = \omega_i^L = \omega_i^R$) and   
\begin{equation} 
S_L = S_R = \sum_j \omega_j \ln \omega_j.   
\label{eq:mutentropy} 
\end{equation}
In what follows this will be denoted by $S$ and called the mutual entropy.
According to Eq.~(\ref{eq:mutentropy}), $S$ is  
the measure of entanglement formation.
In order to describe the relationship between entanglement and mutual entropy 
we can use    
Wootters's interpretation \cite{wootters1} obtained in quantum information theory. 
Within the context of DMRG it can be stated that   
for any pure target state the
entanglement measures the amount of quantum information that must be exchanged between the DMRG  
blocks in order to create the target state. 
 
For a given system block we can generate various 
environment blocks representing "different conditions" and the 
quantum analogue of Eq.~(\ref{eq:kullback}) called Kullback-Leibler entropy or 
quantum relative entropy is written as 
\begin{equation}
K(\rho||\sigma)) \equiv {\rm Tr}(\rho\ln\rho-\rho\ln\sigma),
\label{eq:leibler}
\end{equation}
where $\rho$ and $\sigma$ are two reduced density matrices of the system blocks corresponding to two
different environment blocks.

\subsection{Information generation and annihilation}

In the above treatment a so called static situation was considered where the DMRG system block has a given 
size with given lattice sites (orbitals) and we adjust the configuration space of the environment block. 
During the sweeping procedure of DMRG, however,
the partitioning of lattice sites changes and the system behaves dynamically. In this case for a given 
incident message (given ordering and configuration space of the environment block) the algorithm 
drives the system to an attractor defined by the target state. As it has long been known, 
in a dynamical system, different incident messages can give rise to the same attractor, 
in which case one can speak of redundancy of messages. 
On the other hand, it is possible that one incident message can lead to two different attractors due to the
fluctuation of the system or change of the intrinsic parameter of the system. 
This effect doubles the original information.
As an example, it was found earlier \cite{legeza1} that 
if the symmetry of the target state was not restricted,
then depending on the DMRG parameters the same ordering of lattice 
sites and initial conditions sometimes gave rise to the $S^z_{TOT}=0$ component of the triplet state or 
to the singlet state. 
Clearly the relative importance of the message $\omega_j$ depends not only 
on the dynamical system but also on the tasks it must perform. In order to determine the values of $\omega_j$
of incident messages we have to consider the links between a message and the attractor into which 
the dynamical system is driven after receipt of this message. 
Following again the notation of Haken 
a single message can drive the system via fluctuations into several different attractors
which may occur with branching ratios $M_{jk}$ with $\sum_k M_{jk}=1$.  
Then the relative importance $\omega_j$ is defined as
\begin{equation}
\omega_j = \sum_k L_{jk}\omega_k^{(1)} = \sum_{k}\frac{M_{jk}}{\sum_{j^\prime}M_{j^\prime k} + \epsilon}\omega_k^{(1)}
\end{equation}
where $\epsilon\rightarrow 0$.
If there are several systems coupled one after the other, then 
for instance in a two step procedure we obtain
\begin{equation}
\omega_j = \sum_k L^{(1)}_{jk}\omega_k^{(1)} = \sum_{k_1,k_2} L^{(1)}_{jk_1}L^{(2)}_{k_1 k_2}\omega^{(2)}_{k_2}.
\label{eq:coupled}
\end{equation}
It is worth to emphasize that the recursion from $\omega^{(n)}$ to $\omega$ may depend on the path, 
namely on the ordering of lattice sites or molecule orbitals.
In such a way we obtain an interference of messages and the relative importance of messages
depends on the messages previously delivered to the receiver. In the general case, the 
relative importance of a messages will depend in a non-commuting way on the sequence of 
the messages. In this way the receiver (in our case the system block) is transformed by 
messages again and again and clearly
the relative importance of messages will be a function of the iteration step.

The meaning of Eq.~(\ref{eq:coupled}) is that with
a given task or ensemble of tasks, 
this algorithm allows us to select the message to be sent, namely the one with the biggest $\omega_j$.
Using the relative importance of messages one can investigate the information entropy
of subsequent states of the system
\begin{equation}
S^{(0)} = -\sum_j \omega_j \ln \omega_j,\,\,
S^{(1)} = -\sum_k \omega_k \ln \omega_k,
\end{equation}
whether a dynamical system annihilates, conserves or generates information. 
If $\sum_k \omega_k = 1$ and $\sum_j \omega_j = 1$ there is annihilation of information 
if $S^{(1)}<S^{(0)}$ or generation of information if $S^{(1)}>S^{(0)}$.  
A dynamical system can be "sensitive" 
or "insensitive" for a given message.
As it was described by Haken \cite{haken} it is  
an interesting problem to determine the minimum number of bits required to realize 
a given attractor or to realize a give value of "relative importance". The $M_{min}$
parameter introduced in a previous work \cite{legeza1} had the same meaning. 
It is, however, a major task to determine those incident messages which have the 
largest $\omega_j$ in advance of the calculations.

\section{Numerical investigation of Neumann entropy and entanglement}

The elements of information theory   
outlined in the previous sections have been used in the  
DMRG studies of systems described by the Hamiltonian, 
\begin{equation}
{\cal H} = \sum_{ij\sigma} T_{ij} c^\dagger_{i\sigma}c_{j\sigma} + 
           \sum_{ijkl\sigma\sigma^\prime} {V_{ijkl} 
c^\dagger_{i\sigma}c^\dagger_{j\sigma^{\prime}}c_{k\sigma^{\prime}}c_{l\sigma}},
\label{eq:ham}
\end{equation}
where $T_{ij}$ denotes the matrix elements of the one-particle 
Hamiltonian and $V_{ijkl}$ stands for the matrix elements of the electron 
interaction operator. Depending on the structure of $T_{ij}$ and $V_{ijkl}$ this
Hamiltonian can describe a molecule or a usual fermionic model in solid state physics, e.g.,  
the Hubbard or extended Hubbard models in one or higher dimensions
or for example coupled fermion chains. In the former  
case a one-dimensional chain is built up from the molecular orbitals 
that were obtained, e.g., in a 
suitable mean-field or MCSCF calculation and in the rest of the paper we use 
the numbering of orbitals  
corresponding to the output of these calculations.
In our solid state physics applications the indices $i,j,k,l$  
denote momenta with $k_i = (2\pi n)/L, -L/2< n \leq L/2$.
For the one-dimensional Hubbard model $T_{ij}=-2t\cos(k_i)\delta({i-j})$ and
$V_{ijkl}=(U/L)\delta(i+j-k-l)$. In what follows $U$ is given in units of $t$ with $t=1$. 
In the rest of the paper Hartree-Fock (HF) orbitals denote filled $k_i$ points between 
the Fermi surfaces ($\pm k_F$) in the limit of $V_{ijkl}=0$ where
we use "Fermi surface" to denote sites where the occupation number
of the sites drops to zeros for $V_{ijkl}=0$.   
The full CI (FCI) energy is the exact solution of Eq.~(\ref{eq:ham}) for a finite chain with length $L$ 
and for a given number of electrons with up and down spins. 

\subsection{Analysis on small molecules} 

\begin{figure}
\includegraphics[scale=0.35]{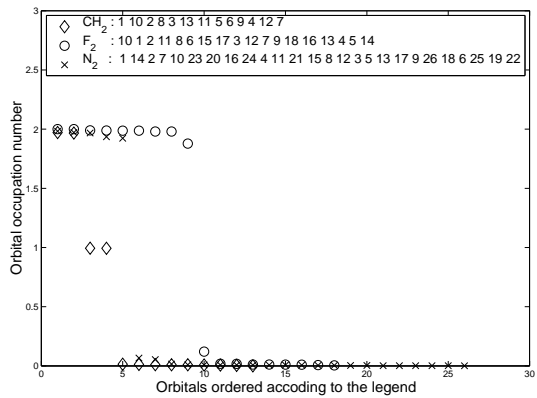}
\vskip .2cm
\caption{Decay of orbital occupation number obtained in the full CI limit and the corresponding 
ordering of orbitals for the molecules studied in the paper.}
\label{fig:occupnums}
\end{figure}

First we have analyzed the density matrices and the energy of the subsystem blocks 
and the interaction between the subsystem blocks by
carrying out test calculations on a very small system, namely the CH$_2$ molecule by correlating 6
electrons on 13 orbitals. The system is so small, that it is hardly
expected that DMRG would not converge to the attractor determined by the target state. 
In the ordering we relied on the  
occupation number of the orbitals obtained in the full CI calculation  
by the MOLPRO program package \cite{MOLPRO}. 
This is plotted in Fig.~\ref{fig:occupnums}
for a few selected test molecules used in the present paper.   
The legend shows the corresponding orbitals with the original indices.
It is worth to mention that for the CH$_2$ molecule 
orbitals $1$ and $10$ are almost doubly occupied while $2$ and $8$ are almost occupied with up or down spins
since the ground state is a triplet state.     

\begin{figure}
\includegraphics[scale=0.45]{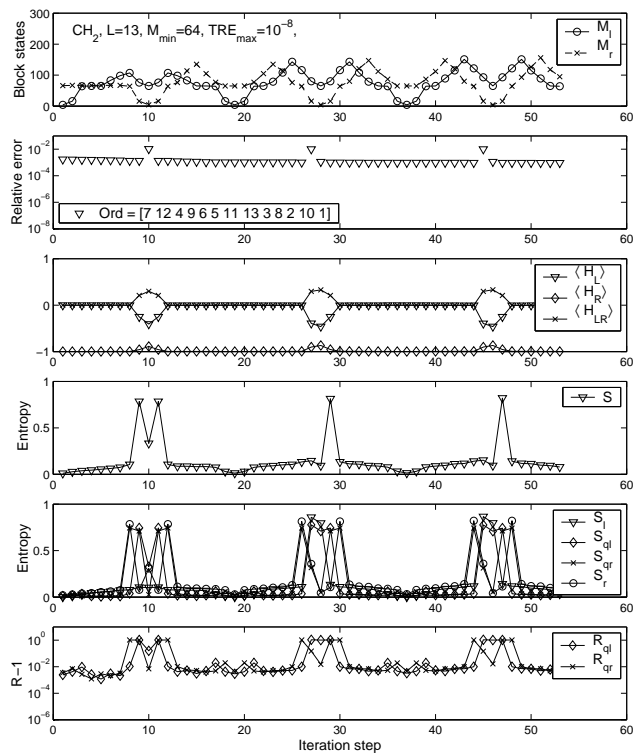}
\caption{Example for an ordering when relevant information of the system is collected into one DMRG block giving rise to vanishing interblock interactions and very low entropy indicating high separability of the target state and lack of quantum information exchange between the blocks.}
\label{fig:ch2_13_ord3}
\end{figure}

In a first attempt 
to generate a good environment block, we have 
put the orbitals with the largest occupation numbers to the right end of the chain. 
This corresponds to the following ordering:  
$[7, 12, 4, 9, 6, 5, 11, 13, 3, 8, 2, 10, 1]$. 
Our result obtained for a truncation error $TRE_{max}=10^{-8}$ is shown in Fig.~\ref{fig:ch2_13_ord3}. 
In subsequent panels we show the number of selected block states, the relative error of the energy,
the relative energy of the blocks and the interaction term, the mutual entropy, 
the site and block entropies and the site participation numbers.  
It is evident from the figure, that
although depending on $TRE_{max}$ the number of selected block states 
fluctuated, the energy 
did not converge to the FCI value. It was trapped at some local minimum.
This extreme situation has often been found in our previous 
studies. Investigating the third panel of the figure, it is obvious that the 
right block alone provides all the contribution to the superblock Hamiltonian as 
$(\langle H_R \rangle \simeq -1)$ and the interaction between the blocks 
vanishes ($\langle H_{LR} \rangle \simeq 0$), except at the turning points. 
Due to the lack of interaction of the blocks the mutual information entropy of the blocks remained
close to zero thus no quantum information exchange was generated during the sweeping procedure as
can be seen on the fourth panel of the figure. 
In order to analyze the information
content of the subsystems we have also plotted in the fifth panel the calculated entropies 
of the left and right blocks 
($S_l$ and $S_r$) and the entropies of the two intermediate sites ($S_{ql}$ and $S_{qr}$).
The sixth panel shows the participation numbers of the two intermediate sites ($R_{ql}$ and $R_{qr}$).
It is clearly seen that sites lying close to the "Fermi
surface" have much larger Neumann entropy ($S_{ql},S_{qr}$) and larger participation number 
($R_{ql},R_{qr}$)  
than orbitals where the occupation number is close to zero or two. (This is determined by identifying
the data points corresponding to the ordering and the partitioning of the composite system for that given
iteration step as was shown in Fig.~\ref{fig:fin01}.) 
The meaning of site entropy and participation number can be explained very easily. 
In our case the maximally mixed state of an orbital described by the $q_l=q_r=4$ 
($0, \uparrow, \downarrow, \uparrow\downarrow$) basis states would correspond to $S=\ln 4$ and $R=4$. 
For those sites
which lie energetically far above ( below) the Fermi surface the $|0\rangle$  
$(|\uparrow\downarrow\rangle)$ basis state in the reduced density matrix of the site would appear with very
large probability and the remaining three basis states with vanishing probability giving rise to $S\simeq0$ 
and $R\simeq 1$. In contrast to this, the reduced density matrices of sites lying close to the Fermi surface 
will have a more uniform eigenvalue spectrum corresponding to a finite value of the entropy and a 
participation number larger than unity. 
Analyzing the situation close to the turning points, we have to recall that 
in this case the $\bullet B_R$ subsystem contains two lattice sites, but orbitals 
$1,10,2,8$ are very important components of the wave function. Therefore, both 
subsystem contain important states and a finite quantum information exchange between the blocks is necessary 
in order to generate the target state. 
As a test we have flipped the ordering 
and found a similar behavior, 
except that the peaks of the interaction terms and the entropies were shifted to 
the other end of the chain.

\begin{figure}
\includegraphics[scale=0.45]{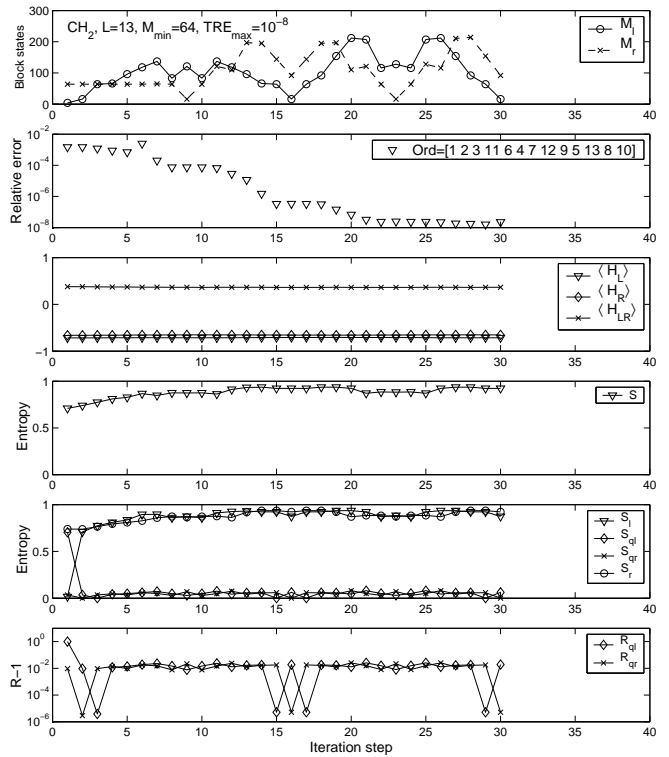}
\caption{Same as Fig.~\ref{fig:ch2_13_ord3} but for an ordering when relevant information 
of the system is shared between DMRG blocks giving rise to strong interblock interaction 
and large quantum information exchange between the blocks.}
\label{fig:ch2_13_ord1}
\end{figure}

As a next step, the interaction between 
the blocks was maximized by putting the orbitals alternatively to the two ends with decreasing occupation number. 
This gives the following ordering:  
$[1, 2, 3, 11, 6, 4, 7, 12, 9, 5, 13, 8, 10]$. 
Our result is shown in
Fig.~\ref{fig:ch2_13_ord1}. It can be seen that the convergence 
became very fast and within one and a half sweeps (26 iterations steps) the error margin 
set by $TRE_{max}$ was reached. This means that the environment error was 
reduced significantly within one full sweep. Investigating the third panel
one can see that there is always a strong interaction between the blocks as
$\langle {\cal H}_{LR}\rangle \simeq 0.5$ and both blocks provide equal amount of energy for the total system as 
$\langle {\cal H}_L\rangle\simeq \langle {\cal H}_R\rangle \simeq -0.7$.  
Analyzing the separability and entanglement of the target state 
the fourth panel shows that the system possesses large mutual entropy indicating a large entanglement and
a large amount of quantum information exchange between the blocks. 
Another major difference compared to the results obtained for the previous ordering is that the
block entropies ($S_l$ and $S_r$) are always very large as can be seen in the fifth panel. 

\begin{figure}
\includegraphics[scale=0.45]{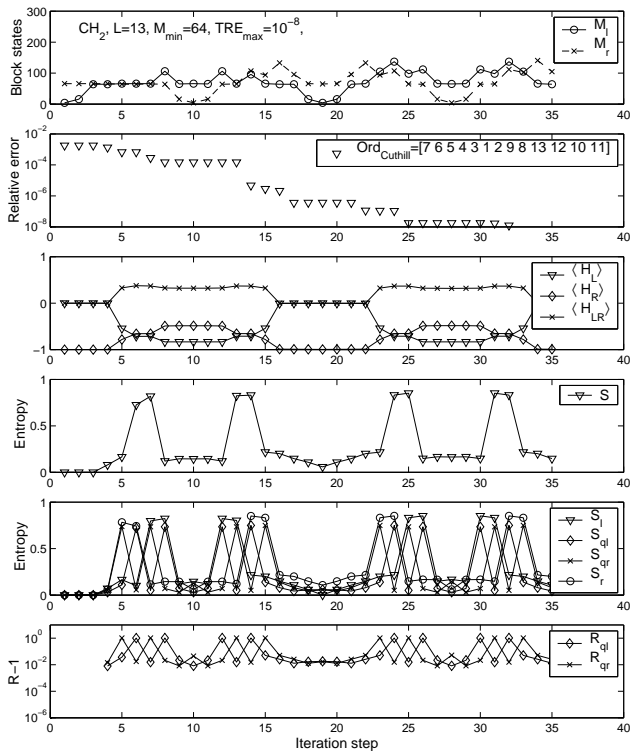}
\caption{Same as Fig.~\ref{fig:ch2_13_ord3} and Fig.~\ref{fig:ch2_13_ord1} but for the Cuthill-McKee ordering.}
\label{fig:ch2_13_ord2}
\end{figure}

Investigating the first panel of Fig.~\ref{fig:ch2_13_ord1} one can see that the maximum number of block states selected 
dynamically were often close to 250. Based
on previous studies we expect that 
the Cuthill-McKee algorithm \cite{cuthill} gives  
a better ordering in the sense that a smaller   
number of block states have to be selected to achieve the same accuracy.
Our result for the ordering $[7, 6, 5, 4, 3, 1, 2, 9, 8, 13, 12, 10, 11]$ is shown in 
Fig.~\ref{fig:ch2_13_ord2}.
It can be seen that in fact the same accuracy was achieved as before with a smaller subset of block states 
($M_{max}<130$). At the same time the mutual entropy had usually large peaks when the energy improved significantly.
Analyzing the fifth and sixth panels one sees that orbitals which have large entropy are now at the 
center of the chain and the site entropies had maxima where the mutual entropy had sharp peaks.
In addition, compared to the two previous orderings it is evident that the minimum of $R$ was
at larger values. 
Looking at the third and fourth panel we can find an extended region with large interblock couplings but 
with small entropy.   
For this region our result seems to indicate smaller entanglement  
between the blocks thus it might correspond  
to a more pronounced finite classical interblock interaction with 
less quantum correlation between the blocks. Again
the significant improvement in the energy happens when the two blocks have the same amount of intrablock energy 
with a large quantum information exchange between the blocks. 
 
\subsection{Analysis on larger molecules} 

\begin{figure}
\includegraphics[scale=0.4]{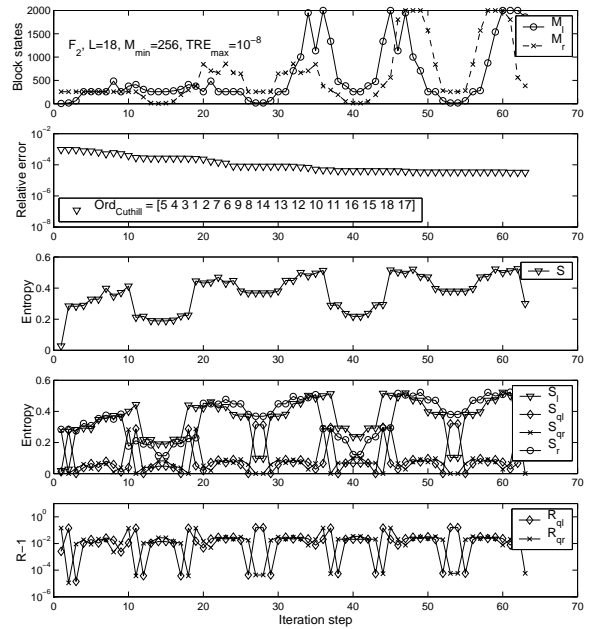}
\caption{Result for F$_2$ molecule correlating 18 electrons on 18 orbitals using Cuthill-McKee ordering.}
\label{fig:f2_18_ordk}
\end{figure}

In order to search for an optimal ordering method 
we studied the quantities defined above 
for the molecules CH$_2$, H$_2$O, F$_2$ and N$_2$.
We have found that in general the Cuthill-McKee ordering 
is not the best one. 
In the case of F$_2$ molecule with 18 electrons on 18 orbitals DMRG did not converge at all,
 as can be seen in Fig.~\ref{fig:f2_18_ordk}.
It is evident from the figure that the orbitals with larger participation number and larger entropy appear 
close to the two ends of the chain and the method performed rather badly. A large number of block states were selected
(we have cut $M_{max}$ at 2000) and the result did not converge.
Cuthill-McKee ordering resulted in a very similar distribution of site entropies as in    
the second example for the CH$_2$ molecule. 
We have tested this kind of entropy distribution  
for the other molecules as well using larger basis sets 
and obtained very slow convergence or lack of convergence. Therefore, the good convergence obtained in the second
example for CH$_2$ molecule was due to the short chain length. 

\subsection{Analysis on the 1-D Hubbard chain} 

\begin{figure}
\includegraphics[scale=0.35]{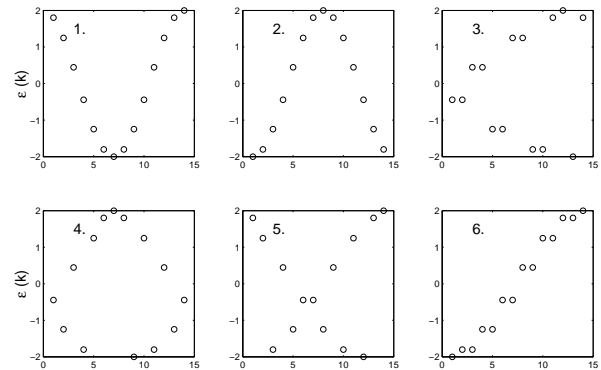}
\vskip .2 cm
\caption{Example for various orderings of $k_i$ points for the half-filled Hubbard chain with L=14. Doubly filled sites in the HF limit in the natural ordering are 4,5,6,7,8,9,10.}
\label{fig:hub_all}
\end{figure}

We have done similar calculations on the 1-D Hubbard model for various band fillings, chain lengths and 
$U$ values. 
Fig.~\ref{fig:hub_all} shows six different orderings of the $k$ values for a half-filled $L=14$ chain.
Using the natural ordering, where the values $-\pi<k_i\leq\pi$ follow each other with
increasing $k_i$, the doubly filled states of the HF state are on the sites 4,5,6,7,8,9,10, the others are empty.
The states at the Fermi energy are on the sites 4,10.

The first panel corresponds to the original ordering, while in the second panel the ordering of sites
with negative and positive momenta were reversed. 
In the third panel sites counted from the Fermi 
surface were ordered to one side of the chain similarly to 
the first example used for CH$_2$. In the fourth ordering the Fermi points were moved to 
the two ends of the chain similarly to the second example used for CH$_2$. The fifth and sixth panels show
our attempts to place the sites near the Fermi surface to the center of the chain as in the third 
example for the CH$_2$ molecule. 

\begin{figure}
\includegraphics[scale=0.4]{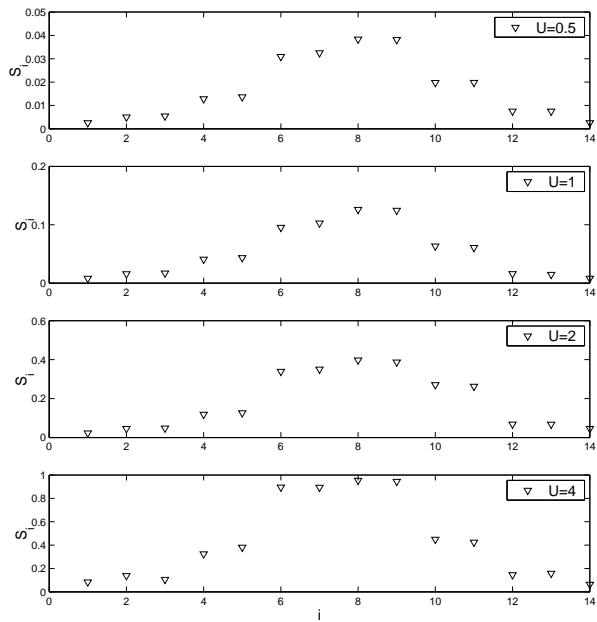}
\vskip .2 cm
\caption{Site entropy ($S_{q}$) obtained for 
the half-filled Hubbard chain with $L=14$ using ordering shown on the 6th panel with U=0.5,1,2,4.}
\label{fig:hub_entropy}
\end{figure}

We have found similar results as for the molecules, namely the third ordering gave the worst result. 
For small $U$ values ($U\simeq 0.5,1$) the method did not converge even if $M_{min}=800$ was used. 
The fourth ordering gave very large mutual entropy and although 
a very large subset
of block states were selected, 
the convergence was still very slow. It required six to seven sweeps
for a half-filled chain with $L=14$, $U=1$, and $TRE_{max}=10^{-6}$. The best performance was
obtained for the fifth and sixth orderings. 
The first two orderings gave stable but considerably slower
convergence. We have carried out several calculations with chain lengths up to 30 sites for various fillings and 
$U$ values and   
found again that good convergence is obtained if the block entropies are larger then a critical value depending
on the various model parameters. 
As an example 
the distribution of site entropy for the half-filled Hubbard model with $L=14$ using the sixth ordering 
is shown in Fig.~\ref{fig:hub_entropy}.
As can be seen larger $U$ values
gave larger values of site entropies and participation numbers which means more mixed states. 
We have also found that the best ordering is 
a function of the filling. This is because the algorithm will converge only if there is 
a finite interaction between the blocks for
several steps, and therefore both blocks must contain sites with large entropies. 
At the same time the reduction of 
block entropies and the number of block states can be achieved by
moving sites with largest entropies close to each other and closer to the center of the chain. 
As an example, for larger
$U=2,4,20$ values, where sites lying farther away from the Fermi surface become more important, we have found 
that DMRG blocks had finite interaction strength for all iteration steps but of course the 
number of maximally selected block states has increased significantly.
As an indication for an $L=14$ half-filled chain a $10^{-5}$ absolute error of the energy has been reached with 
$M_{max}\simeq 400-500$ for $U=0.5$, while the system selected out dynamically more then $2000-2500$ states to
reach an accuracy of $10^{-4}$ for $U=4$. For $L=18$ these numbers are $600-700$ and $3000-3500$, respectively.
This fact questions the efficiency of MS-DMRG when $U$ is comparable to or larger than the band width. We expect that
for the large $U$ limit real space DMRG should provide more accurate results.

\section{Optimizing ordering}

\begin{figure}
\includegraphics[scale=0.4]{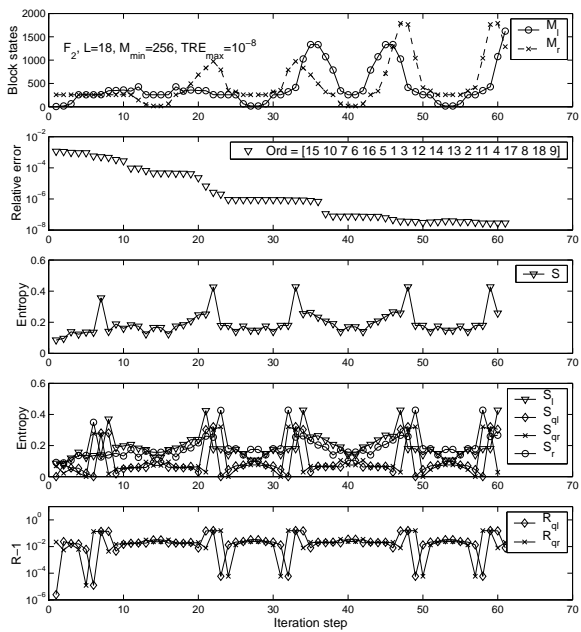}
\caption{Result for F$_2$ molecule correlating 18 electrons on 18 orbitals using our numerically optimized ordering.}
\label{fig:f2_18_ordh_m256}
\end{figure}

It is very important to emphasize that according to Eq.~(\ref{eq:coupled}) the
dynamics of the system depends very much on the path it evolves, thus on the way it receives incident messages
and the renormalized block states are formed. 
In DMRG the initial wave function does not point in the direction of the attractor and it is rotated systematically
in the multidimensional space by the transformation matrices during each renormalization step.
The sequence of rotation, however, is not commuting, therefore, as it has been seen in the 
calculations presented above the ordering has a great impact on the structure of
renormalized states.
For fermionic models 
in solid state physics the optimal ordering
following from our earlier considerations is the one shown on the sixth panel in Fig.~\ref{fig:hub_all}.
It is a far more complicated task to find it in quantum chemistry.
In order to search for an optimal ordering we have  
carried out a kind of brute force calculation for the F$_2$ molecule. 
We have run QC-DMRG using 64 block states for one half sweep, then 
permuted two orbitals and monitored the obtained energy values. If an ordering produced a better result
it was kept and a next ordering was obtained by permuting again two sites.  
For the F$_2$ molecule one calculation took some eight seconds CPU time and more 
than 50 000 permutations were carried out.
Our result for the best ordering is shown in Fig.~\ref{fig:f2_18_ordh_m256}.
It can be seen that the sites with largest entropy are close to the center of the chain
but some sites with finite entropy are close to the two ends of the chain. This configuration is in agreement
with what has been found for the Hubbard chain.  It provides finite interaction 
between the blocks for several steps but the largest information exchange between the blocks occurs when the
superblock composite system is close to a symmetric configuration. 
We have investigated the other test molecules as well and obtained similar results.
It is worth to mention,  
that for a given molecule the ordering which produced the best 
performance at the end of the
first half sweep also gave the fastest convergence using more sweeps.

As a summary of our result, we have obtained the following rule to generate a good ordering, 
which of course could still be not the optimal one: 

{\bf Step 1.} The Cuthill-McKee ordering is used
and a DMRG calculation with some 100 block states is run and the entropies, participation numbers
and orbital occupation numbers are determined.
 
{\bf Step 2.} The orbitals with largest entropies 
(largest participation numbers, closest to the Fermi surface ) are moved to the center
of the chain and the remaining orbitals are distributed along the chain with decreasing entropies  
counted from the two ends of the chain.

This method
guarantees a finite interblock coupling for several steps but also reduces the number of block states.  
It is worth to mention, that  
if $q^r\leq M^{min}_r$  or $q^l\leq M^{min}_l$ then the 
sweeping procedure can be turned back, since the remaining part of the sweeping 
will not improve the environment. All calculations 
in the rest of the paper were obtained following this rule.

\section{The initialization procedure}

Besides ordering, the optimal performance of DMRG is strongly effected by the initial conditions
or in other words by the initial block configurations. 
Looking at Fig.~\ref{fig:ch2_13_ord2} one can see that in  
the first three iteration steps the mutual entropy $S$ was exactly zero,   
 $R_{ql}$ and $R_{qr}$ was unity 
($R-1=0$) while in the second sweep 
at the same partitioning their values became larger indicating a better
environment block. In addition, for the first three iteration steps   
the eigenvalue spectrum of the reduced density matrix of the system block 
($\rho_L$) had one eigenvalue of unity 
which according to  
Eq.~(\ref{eq:sepcond}) means that the superblock eigenstate was separable. For the second sweep all eigenvalues
were less than unity indicating that the target state became non-separable.
Since the subsystem entropies and related quantities depend on the other subsystems, this suggests 
that a better initialization procedure can be obtained if one starts with a larger mixture 
by increasing the block entropies. This also leads to the decrease of the volume 
of separable basis states of the target state. 
We have also found that the entropy of the blocks must be larger than a critical
value.    

The standard initialization procedure 
outlined by Xiang, when subsystem blocks of 
various lengths (describing various partitions) are generated 
in advance of the finite lattice algorithm 
does not guarantee this in general for the first sweep.  
Therefore, it is crucial to develop a new method which  
constructs the environment  
blocks by taking into account the change of the renormalized system block basis states, 
and which increases  the
block entropies above a threshold value during the initialization procedure.  
In this section we present a procedure to generate the system and 
environment blocks giving large $\omega_j$ eigenvalues, i.e., 
block entropies.

\subsection{The DEAS procedure}

Let us consider first the partitioning where the left block (system block) 
contains one lattice site and the right block (environment) $L-3$ lattice sites 
as was shown in Fig.\ref{fig:fin01}. 
The system block contains $M_l=q$ states and the right block contains $M_r$ states 
when the conservation laws of total quantum numbers are taken into account.  
The various  
configurations of the right block interact differently with the left block.   
Since the  
density matrix of the system block depends strongly on the interaction between the two 
blocks the initial configurations of the right 
block have a strong effect on the renormalized block states of the left block. 
In order to guarantee the convergence of DMRG
the mutual entropy should possess a finite value, thus one needs a method  
which maximizes the Kullback-Leibler entropy given by Eq.~(\ref{eq:leibler}), i.e., a protocol to optimize the
right block configuration in advance of the calculation to have largest Neumann entropy.

\begin{figure}
\includegraphics[scale=0.3]{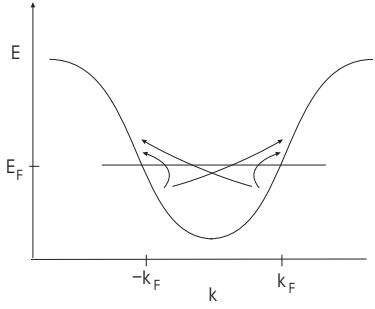}
\caption{Dispersion relation of the 1-D Hubbard chain. Arrows indicate the most probable electron-hole excitations 
with small momentum transfer in the small U limit.} 
\label{fig:disp}
\end{figure}

It is known for the Hubbard
model, that for small $U$  
all the interesting physical properties are determined 
by sites that lie close to the $\pm k_F$ Fermi points. They give also
the largest contribution to the correlation energy. The most important 
excitations are shown in Fig.~\ref{fig:disp}. 
As the strength of U is increased this region gets larger and sites further away from the Fermi points will
also contribute to the correlation energy and other physical properties. We can thus expect that the basis
states formed from the sites around $\pm k_F$ 
will correspond to a larger mixture (several $\omega_j$ will be large) and
we can form the incident message from these states. In order to achieve this we
define a so called AS vector, a concept inherited from quantum chemistry which simply contains 
the most important sites in a descending 
order with respect to their expected importance. For the  
Hubbard chain this vector is $AS\equiv[k_F, -k_F, k_F+1, -k_F-1, k_F-1, -k_F+1, ...]$. 
In QC-DMRG
the AS vector is constructed in a self consistent way.  
First the AS vector is defined to include only the HF orbitals and a quick
calculation is performed with some 100 block states using one or two sweeps. 
Next the AS vector is defined by taking orbitals with  
largest values of site entropies in a descending order. 

We have confirmed numerically that the entropy of the environment block ($S_r$)
can be increased by using sites lying around the Fermi points. 
It is, however, not sufficient to increase $S_r$ only because it could happen that a block configuration with
large $S_r$ could lead to vanishing mutual entropy ($S$) due to the lack of interaction between the blocks.
Once a good AS vector is established, the question arises how to construct the configuration space, i.e., 
how to fill the sites given by the elements of the AS vector.
Therefore, one needs a protocol that  
extends optimally the active space in order to achieve fast convergence.  
In the following we describe the details of the dynamically extended active space (DEAS) procedure which
takes into account the change of the partitioning of the DMRG superblock and 
the renormalization procedure of the system block.

Let us assume that we have a given ordering and we have a starting partitioning with $l=1$ and 
$r=L-2-1$ as is shown in Fig.~\ref{fig:fin01}.

{\bf Step 1.} An AS vector is defined and another vector (HF vector)
is defined which contains the Hartree-Fock orbitals.  

{\bf Step 2.} Based on the ordering the AS and HF vectors are reordered and for the given partitioning  
the elements of the AS and HF vectors belonging to the right block are determined. 

{\bf Step 3.} The $q^3$ bases states of the $B_l \bullet \bullet$ subsystem are formed and their quantum numbers 
are determined.

\begin{figure}
\includegraphics[scale=0.4]{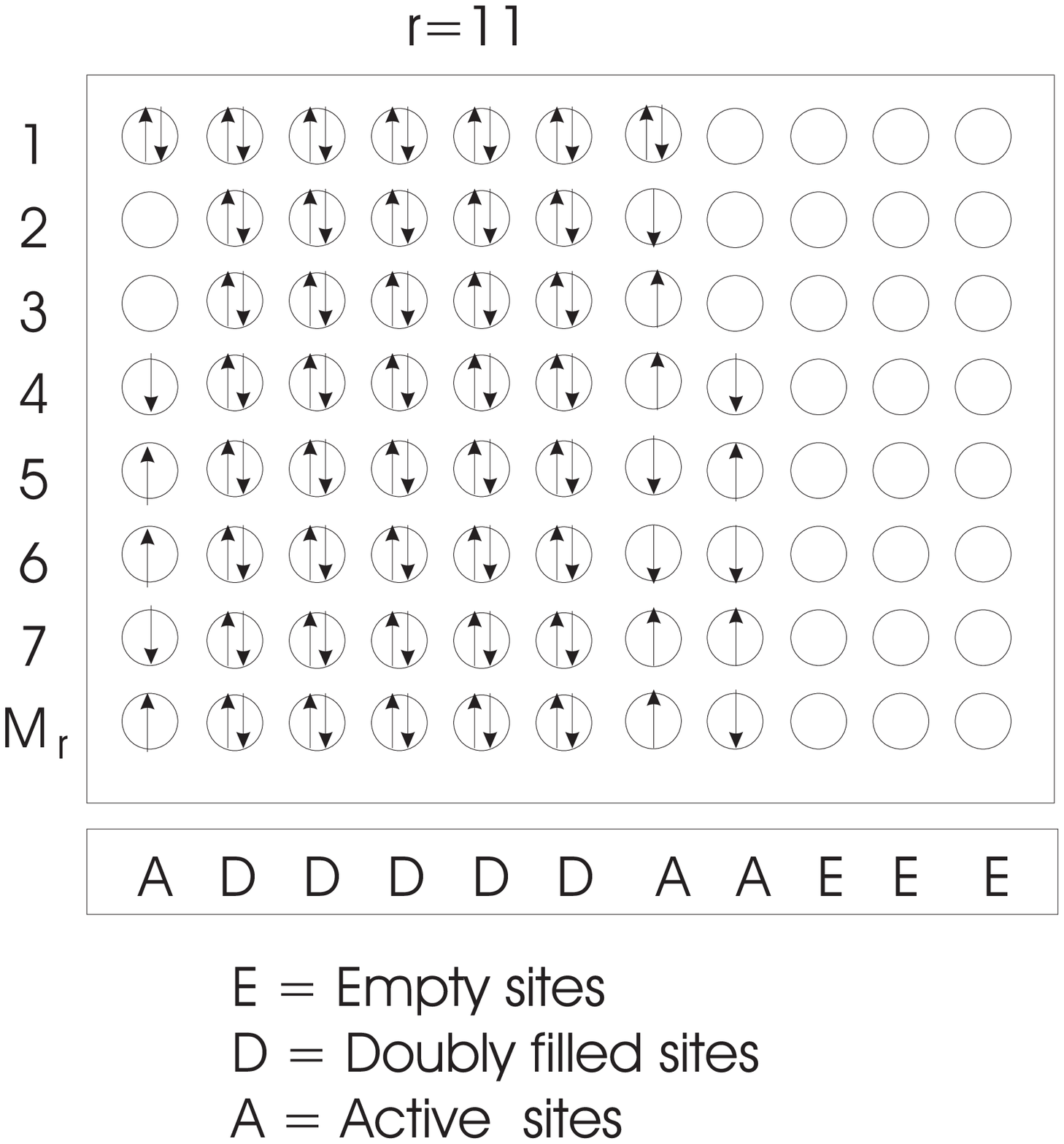}
\vskip .4cm
\caption{Example for a right block configuration space obtained by the DEAS procedure 
for the half-filled Hubbard chain with $L=14$, $N_\uparrow=N_\downarrow=7$, $M_r^{min}=8$, 
$l=1$, $r=11$ and HF vector $\equiv [4,5,6,7,8,9,10]$ and AS vector $\equiv [4,10,3,11,5,9]$.} 
\label{fig:right_block}
\end{figure}

{\bf Step 4.} On the site given by the first element 
of the AS vector the basis states ($0,\uparrow,\downarrow,\uparrow\downarrow$) are formed 
and the HF sites are filled according to the HF configuration except for this site if 
it is in the HF vector. An example is shown in 
Fig.~\ref{fig:right_block} for the half-filled Hubbard chain using the original ordering 
and $M_r^{min}=8$. The
quantum numbers of the obtained states are determined and those which have a matching component with any of 
the $B_l \bullet \bullet$ basis states 
satisfying the conservation of total quantum numbers are kept. Next the first
two elements of the AS vector are taken and the same procedure is repeated 
until $M_r^{min}$ many states are selected. In
this way one makes sure that states of the environment block have a matching component with the system block, thus
they all produce contribution to the mutual entropy.

{\bf Step 5.} One iteration step of the standard DMRG procedure is carried out.   
In the next step then procedures of steps 2-5 are repeated by forming again the $M_l\times q\times q$ states of the 
$B_l \bullet \bullet$ subsystem and the states of the right block matching the left block are determined.  

{\bf Step 6.} Close to the turning points it can happen that for the selected states $M_r<M_r^{min}$ thus the 
excluded states could never 
be recovered during later steps of the method. In order to avoid such problem all $q^r$ states 
are selected for the environment block if $M_r<M_r^{min}$.   

This procedure ensures that the size of the active space is extended dynamically as the system evolves and 
the left block is correlated from the very beginning with the most important states defined by the AS vector.
In addition, as the size of the left and right blocks changes the second step of the procedure adjusts the AS vector
so that only the those elements are taken into account which belong to that given right block. 

\begin{figure}
\includegraphics[scale=0.45]{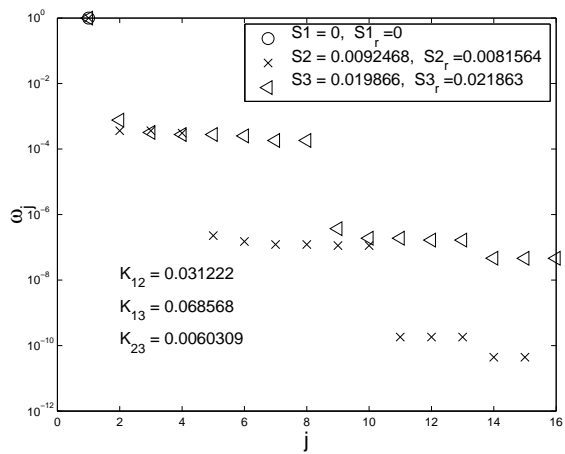}
\vskip .2 cm
\caption{The reduced density matrix eigenvalue spectrum, 
the mutual Neumann entropy (S), environment block entropy ($S_r$) and Kullback-Leibler ($K$) 
entropy calculated for three specially constructed environment blocks 
for the half-filled Hubbard chain with $U=1$ and $L=14$ using the original ordering.}
\label{fig:kullback}
\end{figure}

As an example, Fig.~\ref{fig:kullback} shows the density matrix spectra, the calculated
mutual entropy ($S$), environment block entropy ($S_r$) and Kullback-Leibler entropy ($K$) 
for three specially constructed right blocks with $M_r^{min}=64$ for the half-filled Hubbard 
chain with $U=1$, $L=14$ using the original ordering.  
In the first example we have formed the right block basis states from the sites lying close to the 
right hand side of the chain like in the standard initialization procedure. The second example corresponds to a similar 
construction but the Hartree-Fock sites were doubly filled. 
In the third example the right block was   
set up according to the method  
described above. 
It is evident from the figure, that in the first example one eigenvalue of the 
reduced density matrix was one giving rise to  
zero mutual and environment block entropies and separability of the superblock eigen state.
The second example corresponds to a finite but very small entropy while the entropy is increased
for the third example. 
It is thus clear that a better environment block can be constructed by using the AS vector.

\subsection{Reducing the effective chain length during the initialization procedure}

Once the 
$M_r$ states of the environment block are found, taking 
into account the conservation laws of the quantum numbers, many sites will remain unfilled as
can be seen on the example shown in Fig.~\ref{fig:right_block}. 
If the right block configuration space is represented by an 
$M_r \times r$ matrix then those columns (sites) which contain only the empty state (labeled as E) are not active,   
since the action of creation and annihilation operators on such empty 
sites gives no matrix elements. 
Therefore the 
$i,j,k,l$ summation indices of Eq.~(\ref{eq:ham}) can be restricted only to the remaining non-empty sites (NE) 
giving rise to a much smaller effective size of the chain.

The non-empty sites fall into two further categories. 
Those columns (sites) which contain only doubly filled sites (D) and those called active sites (A) which
contain various basis states $(0,\downarrow,\uparrow,\downarrow\uparrow)$. 
Several steps of the DMRG matrix algebra can be restricted to the active sites only and
various restrictions are also given if an index corresponds to a doubly filled site.
For example the matrix algebra for 
quadratic auxiliary operators having the form $c_{i\alpha}c_{j\beta}$ with $\alpha,\beta$ 
corresponding to up or down spins, or the form 
$c^\dagger_{i\uparrow}c_{j\downarrow}$, have to be carried out only for the active sites.
Therefore, the construction of auxiliary operators \cite{xiang} containing various combinations of one, two, three and four
fermion operators can be obtained in a much faster way. 
This is especially important for the 
quadratic auxiliary operators since their multiplication during the diagonalization of the
superblock Hamiltonian is the most time consuming part of the MS-DMRG. 
It is worth to mention, that the auxiliary operators generated by the left block states 
have be to calculated for the empty sites as well, because during the 
backward sweep the structure of the right block might change and    
non-zero matrix elements in the superblock Hamilton operator could be generated.

This procedure has the advantage,  
that when doubling the number of sites (doing a calculation on a larger basis set), the 
computational time will not increase significantly during the first half sweep and 
a much larger subset of block states (larger $M_r^{min}$)
can be used very efficiently to generate a better environment block. 
For optimal ordering this gives a much faster convergence and 
due to the better environment  
the crossover between the
environment and truncation error can be reached within a few steps. 
After the end of the first half sweep $M_l^{min}=M_r^{min}$ can be used.

\subsection{Application of Abelian point-group symmetry in QC-DMRG}

When the DMRG procedure is applied in quantum chemistry to calculate 
the energy of molecules the use of Abelian point-group symmetry reduces the 
dimension of the Hilbert space to be considered. 
Before proceeding with the numerical results we show how to use this 
symmetry in QC-DMRG.
The irreducible representation of each molecular orbital is generated by  
standard quantum chemistry programs, like the MOLPRO \cite{MOLPRO} program package. The
quantum numbers of the irreducible representations  
are very similar to the momentum quantum numbers $k_i$ used in MS-DMRG. 
The $K_i$ quantum number operator for an orbital defined on the  
$(0,\downarrow,\uparrow,\downarrow\uparrow)$  
basis states in QC-DMRG is written as
\begin{equation}
K_i = k_i  
\begin{array}{ccc}
\left( \begin{array}{c}
1\\
0\\
0\\
1\\
\end{array}\right) & + &
\left( \begin{array}{c}
0\\
1\\
1\\
0\\
\end{array}\right),  
\end{array}  
\end{equation}
where $k_i$ can take values from one to eight according to the symmetry of the orbital. 
The symmetry quantum number operator for basis states of two orbitals is determined by using the  
standard character table as
\begin{equation}
K_{ij} = {\cal T}(K_i,K_j),  
\label{eq:k_symmetry}
\end{equation}
with 
\begin{equation}
{\cal T} = 
\left(\begin{array}{cccccccc}
1 & 2 & 3 & 4 & 5 & 6 & 7 & 8 \\
2 & 1 & 4 & 3 & 6 & 5 & 8 & 7 \\
3 & 4 & 1 & 2 & 7 & 8 & 5 & 6 \\
4 & 3 & 2 & 1 & 8 & 7 & 6 & 5 \\
5 & 6 & 7 & 8 & 1 & 2 & 3 & 4 \\
6 & 5 & 8 & 7 & 2 & 1 & 4 & 3 \\
7 & 8 & 5 & 6 & 3 & 4 & 1 & 2 \\
8 & 7 & 6 & 5 & 4 & 3 & 2 & 1 \,.
\end{array}\right) 
\end{equation}
Using Eq.~(\ref{eq:k_symmetry}) we can determine the symmetry quantum numbers of all the left and right block 
bases states and for a given target state we can restrict the number of electrons and the total symmetry ($K_{TOT}$) 
as 
\begin{equation}
K_{TOT} = {\cal T}(K^L, K^R).
\end{equation}
where $K_{TOT}$ again can take values between one and eight.   
During the renormalization procedure, the symmetry quantum number
operators are renormalized in a similar way as the particle number operators and 
they are assign to the new renormalized block states.

\section{Numerical results using optimized ordering and DEAS}

\begin{figure}
\includegraphics[scale=0.4]{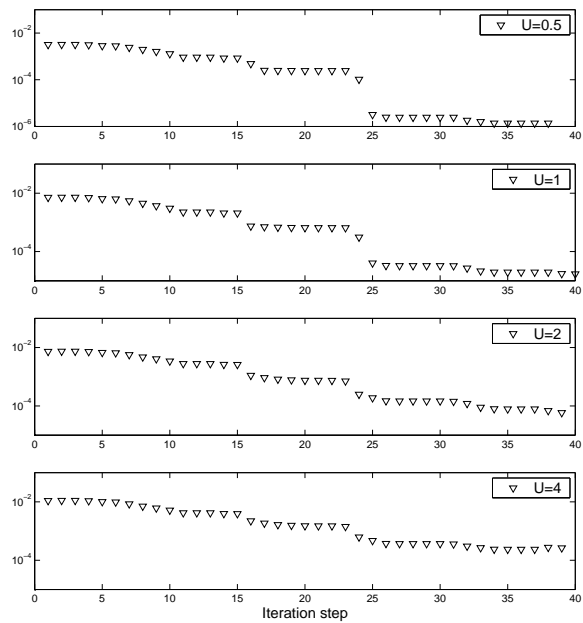}
\vskip .2 cm
\caption{L=18 Half-filled Hubbard chain for $U=0.5,1,2,4$ using DEAS procedure.}
\label{fig:hub_14_all}
\end{figure}

In order to demonstrate the efficiency of DEAS procedure we have studied the 1-D Hubbard chain
for various fillings, orderings, $U$ values and chain lengths. We have found that for small $U$ in the
range $U\simeq 0.1-1$ 
MS-DMRG converges much faster than with the ordinary initialization procedure. 
Performing calculations on half-filled lattices up to 30 sites the error margin set by $TRE_{max}\simeq 10^{-5}$ 
has been reached within one and a half or two sweeps.  
As the strength of $U$ increased the procedure selected out more block states 
but the method still converged much faster. As an example
Fig.~\ref{fig:hub_14_all} shows the result obtained  
for the half-filled, $L=18$ chain for
$U=0.5, 1, 2, 4$. 
In order to obtain an absolute error of $10^{-4}$, 
$TRE_{max}$ was fixed to $10^{-5}$ and  
$M_{min}$ was chosen as $400$ and we used  
the ordering shown in the 6th panel 
of Fig.~\ref{fig:hub_all}. For $U=4$ we have cut the maximum number of block states
at $2500$ giving a maximum value of the truncation error of the order of $10^{-4}$.  
The AS vector is   
$AS\equiv [13,5,14,4,12,6,15,3,11,7,16, ...]$
and the HF vector is $[5,6,7,8,9,10,11,12,13]$.   
It can be seen in the figure, that in all cases the desired accuracy was obtained 
within two sweeps in contrast to the results
obtained by Nishimoto {\sl et al} \cite{erick1} using the standard initialization procedure.
It is even more important to emphasize that after the first (second) sweep the calculated momentum distribution 
$\langle n_k\rangle=\langle \sum_{i,\alpha} c^\dagger_{i\alpha} c_{i\alpha}\rangle$ 
for $U=1 (U=4)$ agreed up to 3 digits with
the values obtained by real space DMRG with periodic boundary condition.

\begin{figure}
\includegraphics[scale=0.45]{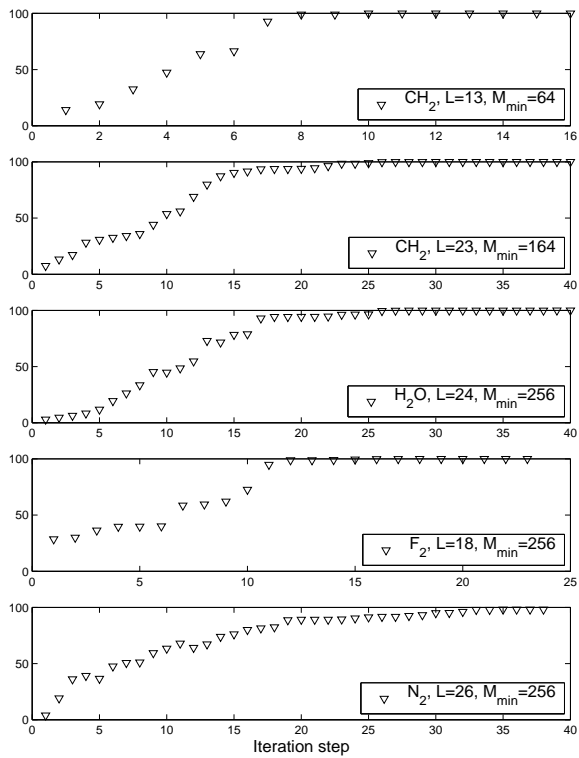}
\vskip .5cm
\caption{Percent of correlation energy as a function of iteration step obtained by DEAS procedure for the CH$_2$, H$_2$O, N$_2$, F$_2$ molecules.}
\label{fig:cas_all}
\end{figure}

Similar results were found for the molecules studied. Results for the CH$_2$ molecule for 
fillings (number of electrons/number of orbitals) 6/13 and 6/23, H$_2$O for 8/24, 
N$_2$ for 10/26 and F$_2$ for 18/18 are shown in 
Fig.~\ref{fig:cas_all}.
In all cases the AS vector was determined in a 
self-consistent way by first running a quick full sweep
with some 64 block states and then it was constructed based on the decay of the site entropy values  
described in the previous section. In all cases we have used the modified Cuthill-McKee ordering,  
according to the method described in Sec.~ IV. 
It is clear from the figure that the DMRG with DEAS procedure converged very fast and
usually 90-99 percent of the correlation energy was obtained within the first one half sweep. 
Therefore, the crossover between the environment and the truncation error can be reached much faster 
and the accuracy of the method is finally determined by $TRE_{max}$ as was shown in Ref.~[\onlinecite{legeza1}].
As an example for CH$_2$ for 6/13 filling the same accuracy that was shown in Fig.~\ref{fig:ch2_13_ord2} was reached in 13 
iteration steps with DEAS procedure using the same parameter sets. For F$_2$ the same accuracy that was 
shown in Fig.~\ref{fig:f2_18_ordh_m256}
was reached in 23 iteration steps.

It is very important to mention, that besides the very fast convergence, 
if an optimal ordering is obtained and a good initial condition is used then the
number of block states can be reduced significantly during the process of renormalization and finally a block can
describe the total system very accurately. 
As an example we have set $M_{min}=4$ and 
the energy of the CH$_2$ molecule with 13 orbitals was 
obtained after the second sweep up to 7 digits accuracy with $M_l=21, M_r=4, \dim {\cal G} = 107$ 
and for F$_2$ 18/18 up to 5 digits accuracy with $M_l=120, M_r=4, \dim {\cal G} = 220$ 
in contrast to the dimension of the exact solution which are 230230 and 9075135300 calculated respectively by
the ${(2L)!}/{((2L-N)!N!)}$ formula. The same was obtained for the half-filled Hubbard model up to 4 digits 
accuracy with
 $L=18, U=1/2$ is $M_l=280, M_r=4 \dim {\cal G}=562$.  
This feature is expected to have a close relationship to 
quantum data compression \cite{schumacher1,jozsa1}. 
Therefore, it seems to be promising to examine 
the separability of reduced density matrices
and their effect on the number of block states. A more detailed study of the relationship between site entropies
 and the dimension
of the Hilbert space of the superblock Hamiltonian based on Suchmacher quantum data compression is in progress.

\section{Summary and future prospectives}

We have studied 
the effect of ordering of lattice sites in 
the momentum space version of the 
density matrix renormalization group (MS-DMRG) method by solving the 1-D Hubbard chain and various molecules.
This was done by calculating site and block entropies, the separability and entanglement of the target state.
Our findings are listed below: 

(1) We have shown that the $k_i$ sites or molecular orbitals lying closer
to the Fermi surface have larger information content and correspond to a larger mixture described 
by larger Neumann entropy and participation number.

(2) We have shown that if all the sites with large entropy are placed close to one end of the chain then 
this improper ordering can lead to vanishing mutual entropy,  
lack of entanglement and quantum information exchange between the system block and environment block.
This prevents the method 
to convergence to the target state. Therefore, the proposal by Xiang to place the highly correlated sites
as close as possible is not a sufficient condition alone.   

(3) If lattice sites with 
large entropy are moved to the two ends of the chain the block entropies can be maximized
giving rise to a larger entanglement, but the convergence of the method can still be very slow or
the method does not converge at all. Numerical
results indicate that if these sites are moved to the center of the chain then a very fast convergence can be 
obtained with much smaller subset of block states. We have found that the mutual entropy has sharp peaks when 
the most active sites are just between the system and environment blocks when the
energy also improves significantly.  Furthermore, in order to have a finite interaction 
between the blocks some sites with smaller 
but finite entropies 
have to be placed close to the ends of the chain. This keeps the block entropy above a critical value 
for several iteration steps.   
We have shown that these conditions are often not satisfied when  
using the Cuthill-McKee algorithm.      

(4) We have developed a new initialization procedure which gives finite block entropies even during the first
half sweep of MS-DMRG and reduces the volume of separable basis states of the target state. This new method
extends the active space in a dynamical fashion 
resulting in a very accurate set of renormalized block states and a very fast convergence. 
For the models that have been studied 90-99 percent of the correlation energy was obtained within the
first half sweep.

(5) We have shown that MS-DMRG is very sensitive to the ordering and the initial conditions.  
A good starting configuration and an optimized ordering can result in  
very accurate blocks with very limited number of block states which still describes the system with
the accuracy determined in advance.

We expect that DMRG can be a very good candidate for a 
new method on quantum data compression and on quantum error correction \cite{qecc1,qecc2,qecc3}
in which case the fidelity can be 
defined in advance by $TRE_{max}$. This research is in progress.

Since MS-DMRG describes a composite system with long-range interactions it is expected that 
the method could be used 
in the context of non-extensive thermodynamics 
to study models
for which the subadditivity of the subsystem entropy does not hold. This would allow 
one to investigate
Tsallis entropy \cite{tsallis}, non-extensive mutual entropy
and quantum entanglement by using the generalized Kullback-Leibler entropy \cite{abe2}.

\acknowledgments
This research was supported in part by the Fonds der Chemischen Industrie and   
the Hungarian Research Fund(OTKA) Grant No.\ 30173,  32231 and 43330. \"O.~L.~acknowledges
useful discussions with B.~A.~Hess and G.~F\'ath. The authors also thank 
Holger Benthien for providing 
results for the Fermi momentum distribution curve obtained with real space DMRG using periodic 
boundary conditions.


\end{document}